\documentclass[seceq,preprint]{ptptex}

\usepackage{graphicx}
\usepackage{feynmf}

\newcommand{\Dslash}{{\not}\kern-0.05em D}
\newcommand{\dslash}{{\not}\kern+0.1em\partial}
\newcommand{\nablaslash}{{\not}\kern+0.05em\nabla}

\newcommand{\tr}{\mathop{\rm tr}\nolimits}

\newcommand{\SU}{\mathop{\rm SU}}
\newcommand{\SO}{\mathop{\rm SO}}
\newcommand{\U}{\mathop{\rm {}U}}
\newcommand{\rmd}{{\rm d}}

\newcommand\fverb{\setbox\pippobox=\hbox\bgroup\verb}
\newcommand\fverbdo{\egroup\medskip\noindent%
                        \fbox{\unhbox\pippobox}\ }
\newcommand\fverbit{\egroup\item[\fbox{\unhbox\pippobox}]}
\newbox\pippobox

\preprintnumber[3cm]{
RIKEN-TH-76\\DPNU-06-08\\YITP-06-44
\\{\tt hep-th/0609049}
}

\title{
Note on Massless Bosonic States in Two-Dimensional Field Theories%
}

\author{
Hidenori \textsc{Fukaya},$^{1,}$\footnote{E-mail: hfukaya@riken.jp}
Masashi \textsc{Hayakawa},$^{2,}$\footnote{E-mail:
hayakawa@eken.phys.nagoya-u.ac.jp}
Issaku \textsc{Kanamori},$^{3,}$\footnote{E-mail: kanamori-i@riken.jp}\\
Hiroshi \textsc{Suzuki},$^{4,}$\footnote{E-mail: hsuzuki@riken.jp}
and
Tomohisa \textsc{Takimi},$^{5,}$\footnote{E-mail: ttakimi@riken.jp}
}

\inst{
$^{1,3,4,5}$Theoretical Physics Laboratory, RIKEN, Wako 351-0198,
Japan\\
$^2$Department of Physics, Nagoya University, Nagoya 464-8602, Japan\\
$^5$Yukawa Institute for Theoretical Physics, Kyoto University, Kyoto 606-8502,
Japan\\
}

\abst{
In a wide class of $G_L\times G_R$ invariant two-dimensional
super-renormalizable field theories, the parity-odd part of the two-point
function of global currents is completely determined by a fermion one-loop
diagram. For any non-trivial fermion content, the two-point function possesses
a massless pole which corresponds to massless bosonic physical states. As an
application, we show that two-dimensional $\mathcal{N}=(2,2)$ supersymmetric
gauge theory without a superpotential possesses $\U(1)_L\times\U(1)_R$ symmetry
and contains one massless bosonic state per fixed spatial momentum. The
$\mathcal{N}=(4,4)$ supersymmetric pure Yang-Mills theory possesses
$\SU(2)_L\times\SU(2)_R$ symmetry, and there exist at least three massless
bosonic states.
}

\begin{document}

\maketitle

\section{Introduction}
In this paper, we show that in a wide class of $G_L\times G_R$ invariant
two-dimensional super-renormalizable field theories, the parity-odd part of the
two-point function of global currents can be determined to all orders in
perturbation theory. The two-point function possesses a massless pole for any
non-trivial fermion content, and this fact provides a simple criterion to
determine the existence of massless bosonic physical states without solving the
dynamics. Our argument is based on anomalous Ward-Takahashi (WT)
identities and is somewhat similar to that for the 't~Hooft anomaly matching
condition.\cite{'tHooft:1979bh,Frishman:1980dq,Coleman:1982yg} \ In fact,
applying the anomaly matching argument to the systems we consider
(assuming that the anomalous behavior of the two-point function does not
receive higher-order radiative corrections), one would arrive at a similar
conclusion concerning massless states. (See, for example,
Refs.~\citen{Elitzur:1981gh,Craigie:1984ah,Witten:1995im}.) The point of this
paper is, however, to show that in the two-dimensional systems we consider, an
elementary argument suffices to obtain an explicit form of the two-point
function to all orders in perturbation theory.
In particular, our argument is applicable even to systems in which the left
and right moving modes are not decoupled due to the Yukawa interaction and the
scalar potential.\footnote{For such a system, application of the standard
technique of conformal field theory is not straightforward.} This point is
crucial for application to two-dimensional extended supersymmetric gauge
theory, which is our main concern.
As an illustration, we show that all solvable super-renormalizable theories in
which the existence of massless bosonic states is known are covered by
the argument. (For earlier studies on these two-dimensional field theories,
see Ref.~\citen{Abdalla:1991vu}.)
We then consider two-dimensional supersymmetric gauge
theories and observe that the $\mathcal{N}=(2,2)$ supersymmetric gauge theory
without a superpotential and the $\mathcal{N}=(4,4)$ supersymmetric pure
Yang-Mills theory contain massless bosonic states. We can also determine
the explicit form of the two-point function of global currents in these
theories. Combined with supersymmetric WT identities, these findings should be
useful in examining recently developed lattice formulations of these
two-dimensional supersymmetric gauge
theories.\cite{Kaplan:2002wv}\tocite{Endres:2006ic}

\section{All-order proof of the massless bosonic state}
We consider two-dimensional field theories of the form\footnote{Throughout the
paper, we use the following notational
conventions. The Greek indices $\mu$, $\nu$, \dots\ run from 0 to~1. The flat
metric is of lorentzian signature, $g_{\mu\nu}=\mathop{\rm diag}(1,-1)$.
We have
$\{\gamma^\mu,\gamma^\nu\}=2g^{\mu\nu}$, $\overline\psi=\psi^\dagger\gamma_0$,
$\gamma_5=\gamma^0\gamma^1$ and~$\mathcal{P}_\pm=(1\pm\gamma_5)/2$. The
chiralities are defined by $\psi_{R,L}=\mathcal{P}_\pm\psi$ and
$\overline\psi_{R,L}=\overline\psi\mathcal{P}_\mp$. The anti-symmetric
tensor~$\epsilon^{\mu\nu}=-\epsilon^{\nu\mu}$ is defined by $\epsilon^{01}=+1$.
}
\begin{equation}
   \mathcal{L}=-{1\over4}F_{\mu\nu}^AF^{A\mu\nu}
   +\overline\psi i\gamma^\mu D_\mu\psi
   +{1\over2}D_\mu\sigma D^\mu\sigma+{1\over2}D_\mu\pi D^\mu\pi
   -V(\sigma,\pi)-Y(\psi,\overline\psi,\sigma,\pi),
\label{twoxone}
\end{equation}
where $F_{\mu\nu}^A=
\partial_\mu A_\mu^A-\partial_\mu A_\mu^A+gf_{ABC}A_\mu^BA_\nu^C$ are the
field strengths of the gauge fields~$A_\mu^A$ (with $f_{ABC}$
are the structure constants of the gauge group), and
$D_\mu=\partial_\mu-igA_\mu^At^A$ is the gauge covariant derivative. The
representation~$t^A$ may differ for each field, and the fermion belongs to the
representation~$r$ of the gauge group.
The potential energy, $V$, and the Yukawa
interaction, $Y$, contain scalar fields $\sigma$ and~$\pi$ (but not their
derivatives), and $Y$ is bi-linear in the fermion fields. We assume that the
system
possesses parity invariance with the assignment that $\sigma$ is a scalar and
$\pi$ is a pseudo-scalar. Power counting in two dimensions, we find that all
coupling constants contained in $D_\mu$, $V$ and~$Y$ are dimensional and thus
the system is super-renormalizable.

We assume that the lagrangian density~$\mathcal{L}$ in Eq.~(\ref{twoxone})
possesses $G_L\times G_R$ global flavor symmetry. That is, $\mathcal{L}$ is
invariant under the transformations
\begin{equation}
   \psi_L(x)\to\exp\{i\theta_L^a T^a\}\psi_L(x),
   \qquad\psi_R(x)\to\psi_R(x),
\label{twoxtwo}
\end{equation}
and
\begin{equation}
   \psi_R(x)\to\exp\{i\theta_R^a T^a\}\psi_R(x),
   \qquad\psi_L(x)\to\psi_L(x),
\label{twoxthree}
\end{equation}
where $\theta_L^a$ and $\theta_R^a$ are independent global parameters, and
$T^a$ are hermitian generators of a certain compact Lie group~$G$ in the
representation~$R$, if supplemented with suitable transformations of the scalar
fields $\sigma$ and~$\pi$. In the following discussion, it is very important
that the generators~$T^a$ in Eqs.~(\ref{twoxtwo}) and~(\ref{twoxthree}) are
the same. More precisely, they are in the {\it same\/} representation~$R$ of
the group~$G$, and we use the terminology $G_L\times G_R$ in this restricted
sense.
This property depends on the structure of the Yukawa interaction~$Y$ and the
scalar potential~$V$ whose explicit forms we do not specify.

For the quantization of the system~(\ref{twoxone}), we need gauge fixing and
introduction of Faddeev-Popov (FP) ghosts. Although we do not explicitly write
down these additional parts, we assume that the gauge fixing condition respects
the Lorentz invariance and global symmetries of~$\mathcal{L}$.

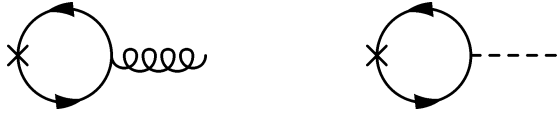
\begin{figure}
\centerline{
\unitlength=0.5mm
\begin{fmffile}{Figure1}
   \begin{fmfgraph}(50,30)\fmfpen{thin}
   \fmfleft{v1}
   \fmfright{v2}
   \fmfv{decor.shape=cross,decor.size=5thick}{v1}
   \fmf{plain_arrow,right}{v1,v3,v1}
   \fmf{curly,tension=2.0}{v3,v2}
   \end{fmfgraph}
\hskip20mm
   \begin{fmfgraph}(50,30)\fmfpen{thin}
   \fmfleft{v1}
   \fmfright{v2}
   \fmfv{decor.shape=cross,decor.size=5thick}{v1}
   \fmf{plain_arrow,right}{v1,v3,v1}
   \fmf{dashes,tension=2.0}{v3,v2}
   \end{fmfgraph}
\end{fmffile}
}
\caption{Some parts of Feynman diagrams which contain the global
currents~$J_{L,R,\mu}^a$. The circle, curly line and dashed line represent
a fermion, gauge boson and scalar boson, respectively, and the cross represents
current operators. We assume that the $G_L\times G_R$ symmetry of $\mathcal{L}$
does not suffer from the anomaly in the usual sense. Specifically, we assume
that the contraction of these diagrams with $-ip_\mu$, where $p_\mu$ is the
momentum flowing into the current operator, vanishes.}
\label{fig1}
\end{figure}

\begin{figure}
\centerline{
\unitlength=0.5mm
\begin{fmffile}{Figure2}
   \begin{fmfgraph}(50,30)\fmfpen{thin}
   \fmfleft{v1}
   \fmfright{v2}
   \fmfv{decor.shape=cross,decor.size=5thick}{v1,v3}
   \fmf{plain_arrow,right}{v1,v3,v1}
   \fmf{curly,tension=2.0}{v3,v2}
   \end{fmfgraph}
\end{fmffile}
}
\caption{If the $G_L\times G_R$ symmetry is gauged, this diagram gives rise to
the breaking of the current conservation.}
\label{fig2}
\end{figure}
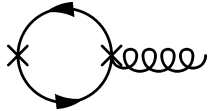

Now, corresponding to the global symmetry $G_L\times G_R$ of
Eqs.~(\ref{twoxtwo}) and~(\ref{twoxthree}), there exist the Noether currents
$J_{L,\mu}^a(x)$ and~$J_{R\mu}^a(x)$, respectively. We assume that the above
symmetry is strictly global, i.e., that there are no gauge fields which couple
to the Noether currents. We also assume that the current conservation of these
Noether currents does not suffer from the anomaly in the usual sense.
Specifically,
we assume that the divergence of the diagrams in Fig.~\ref{fig1} identically
vanishes.
This assumption would not hold if the $G_L\times G_R$ symmetry were gauged,
because the diagram in Fig.~\ref{fig2} leads to the breaking of the
current conservation.


For the following treatment, it is convenient to define linear combinations of
$J_{L,R,\mu}^a$ as
\begin{equation}
   J_\mu^a=J_{R,\mu}^a+J_{L,\mu}^a,\qquad
   J_{5\mu}^a=J_{R,\mu}^a-J_{L,\mu}^a,
\label{twoxfour}
\end{equation}
taking the forms
\begin{equation}
   J_\mu^a(x)=\overline\psi\gamma_\mu T^a\psi(x)+\cdots,\qquad
   J_{5\mu}^a(x)=\overline\psi\gamma_\mu\gamma_5T^a\psi(x)
   +\cdots,
\label{twoxfive}
\end{equation}
where omitted terms are possible contributions from scalar
fields.\footnote{If there is no Yukawa interaction (i.e., $Y=0$) as is the case
in two-dimensional massless QCD, there are no omitted terms in
Eq.~(\ref{twoxfive}), and one may require the condition
$\epsilon_{\mu\nu}J^{a\nu}=-J_{5\mu}^a$ on current operators, as suggested
by the identity~$\epsilon_{\mu\nu}\gamma^\nu=-\gamma_\mu\gamma_5$. Then
$J_L^{a0}+J_L^{a1}=J_R^{a0}-J_R^{a1}=0$, and the current conservation condition
$\partial_\mu J_{L,R}^{a\mu}=0$ implies that the combinations
$J^{a-}\equiv J_L^{a0}-J_L^{a1}$ and
$J^{a+}\equiv J_R^{a0}+J_R^{a1}$ are chiral in the sense that
$(\partial_0-\partial_1)J^{a-}=0$ and
$(\partial_0+\partial_1)J^{a+}=0$ (i.e.,
$J^{a-}$ and $J^{a+}$ are left- and right-moving, respectively).
The current algebra is thus decomposed into left- and right-moving Kac-Moody
algebras.\cite{Craigie:1984ah} \ This decomposition, however, cannot carried
out if there is a non-zero Yukawa interaction.}
We now consider the two-point function of the global currents
$\langle0|T^*J_{5\mu}^a(x)J_\nu^b(y)|0\rangle$. We show below that, under
the assumptions placed on the two-dimensional model~(\ref{twoxone}), this
two-point function receives a contribution only from a fermion one-loop
diagram and can be determined to all orders in perturbation theory.

The most general form of the two-point function that is consistent with
Lorentz covariance and parity invariance is given by\footnote{Note the
identity $p_\mu\epsilon_{\nu\rho}p^\rho-p_\nu\epsilon_{\mu\rho}p^\rho=
-p^2\epsilon_{\mu\nu}$.}
\begin{equation}
   \mathop{\rm FT}\langle0|
   T^*J_{5\mu}^a(x)J_\nu^b(y)|0\rangle
   =-{i\over2\pi}
   \left\{
   {1\over p^2}F^{ab}(p^2)(p_\mu\epsilon_{\nu\rho}p^\rho
   +p_\nu\epsilon_{\mu\rho}p^\rho)+G^{ab}(p^2)\epsilon_{\mu\nu}
   \right\},
\label{twoxsix}
\end{equation}
where ``$\mathop{\rm FT}$'' denotes the Fourier transformation with
$\int\rmd^2x\,e^{ip(x-y)}$. The divergence of the two-point function yields
\begin{eqnarray}
   &&\mathop{\rm FT}
   \langle0|T^*\partial_\mu J_5^{a\mu}(x)J_\nu^b(y)|0\rangle
   =-{1\over2\pi}\{F^{ab}(p^2)-G^{ab}(p^2)\}\epsilon_{\nu\rho}p^\rho,
\nonumber\\
   &&\mathop{\rm FT}
   \langle0|T^*J_{5\mu}^a(x)\partial_\nu J^{b\nu}(y)|0\rangle
   ={1\over2\pi}\{F^{ab}(p^2)+G^{ab}(p^2)\}\epsilon_{\mu\rho}p^\rho.  
\label{twoxseven}
\end{eqnarray}

Next we note that naive WT identities based on the $G_L\times G_R$ invariance
would imply that the quantities in Eq.~(\ref{twoxseven}) are equal to
$if_{abc}\langle0|J_{5\nu}^c(0)|0\rangle$
and $-if_{abc}\langle0|J_{5\mu}^c(0)|0\rangle$, respectively
(where $f_{abc}$ are the structure constants of the group~$G$), which vanish
due to Lorentz invariance of the vacuum. This implies that $F^{ab}=G^{ab}=0$
and Eq.~(\ref{twoxsix}) identically vanishes. Thus, the two-point
function~(\ref{twoxsix}) can be non-zero only as a result of an anomalous
breaking of the WT identities. This ``anomaly'' (which is exactly the central
extension of the current algebra in two dimensions) can arise only from
potentially UV
divergent diagrams, because the naive WT identities safely apply to UV
convergent diagrams. The point here is that under the above assumptions placed
on the model~(\ref{twoxone}), it turns out that only a fermion one-loop
diagram contributes to Eq.~(\ref{twoxseven}). Thus, to all orders in
perturbation theory, the coefficients $F^{ab}$ and~$G^{ab}$ in
Eq.~(\ref{twoxsix}) can be completely determined.

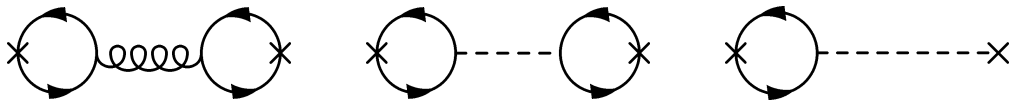
\begin{figure}
\centerline{
\unitlength=0.5mm
\begin{fmffile}{Figure3}
   \begin{fmfgraph}(70,30)\fmfpen{thin}
   \fmfleft{v1}
   \fmfright{v2}
   \fmfv{decor.shape=cross,decor.size=5thick}{v1,v2}
   \fmf{plain_arrow,right}{v1,v3,v1}
   \fmf{plain_arrow,right}{v2,v4,v2}
   \fmf{curly,tension=1.5}{v3,v4}
   \end{fmfgraph}
\hskip10mm
   \begin{fmfgraph}(70,30)\fmfpen{thin}
   \fmfleft{v1}
   \fmfright{v2}
   \fmfv{decor.shape=cross,decor.size=5thick}{v1,v2}
   \fmf{plain_arrow,right}{v1,v3,v1}
   \fmf{plain_arrow,right}{v2,v4,v2}
   \fmf{dashes,tension=1.5}{v3,v4}
   \end{fmfgraph}
\hskip10mm
   \begin{fmfgraph}(70,30)\fmfpen{thin}
   \fmfleft{v1}
   \fmfright{v2}
   \fmfv{decor.shape=cross,decor.size=5thick}{v1,v2}
   \fmf{plain_arrow,right}{v1,v3,v1}
   \fmf{dashes,tension=0.9}{v3,v2}
   \end{fmfgraph}
\end{fmffile}
}
\caption{Example of UV divergent diagrams in which the UV divergence arises
from a loop containing only a single current operator.}
\label{fig3}
\end{figure}

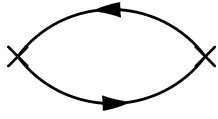
\begin{figure}
\centerline{
\unitlength=0.5mm
\begin{fmffile}{Figure4}
   \begin{fmfgraph}(50,30)\fmfpen{thin}
   \fmfleft{v1}
   \fmfright{v2}
   \fmfv{decor.shape=cross,decor.size=5thick}{v1,v2}
   \fmf{plain_arrow,right=0.5}{v1,v2}
   \fmf{plain_arrow,right=0.5}{v2,v1}
   \end{fmfgraph}
\end{fmffile}
}
\caption{The fermion one-loop diagram which contributes to the
divergences~(\ref{twoxseven}).}
\label{fig4}
\end{figure}

Since our model~(\ref{twoxone}) is super-renormalizable, among all Feynman
diagrams that contain two global currents, only a few are UV
divergent. Let us enumerate such UV divergent diagrams. The first such class
comprises
diagrams such as those shown in Fig.~\ref{fig3}; they diverge at a loop which
contains only a single current. This class, however, does not contribute to
Eq.~(\ref{twoxseven}), because of our assumption of current
conservation for the Feynman diagrams appearing in Fig.~\ref{fig1}.
Therefore the diagrams that contribute to the non-conservation of current
described by Eq.~(\ref{twoxseven}) are those in which the UV divergence arises
from a loop which contains both currents $J_{5\mu}^a$ and $J_\nu^b$. We note,
however, that $J_{5\mu}^a$ contains an odd number of $\pi$~fields and $J_\nu^b$
contains an even number of~$\pi$~fields, due to the parity invariance.
Therefore, there is no scalar one-loop diagram that contains both currents,
because the scalar propagator connects only $\sigma$-$\sigma$ and $\pi$-$\pi$
pairs. In this way, we conclude that the unique diagram which contributes to
Eq.~(\ref{twoxseven}) is the fermion one-loop diagram of Fig.~\ref{fig4}.

Applying a standard argument\cite{Adler} to the diagram in Fig.~\ref{fig4},
we have
\begin{eqnarray}
   &&\mathop{\rm FT}\langle0|
   T^*\overline\psi\gamma_\mu\gamma_5T^a\psi(x)\,
   \overline\psi\gamma_\nu T^b\psi(y)|0\rangle
\nonumber\\
   &&=-{i\over2\pi}(\dim r)T(R)\delta_{ab}\left\{
   {1\over p^2}(p_\mu\epsilon_{\nu\rho}p^\rho+p_\nu\epsilon_{\mu\rho}p^\rho)
   +L(p^2)\epsilon_{\mu\nu}\right\},
\label{twoxeight}
\end{eqnarray}
where $\dim r$ is the dimension of the gauge-group representation~$r$, and
$T(R)$ is the second Casimir of the ``flavor-group'' representation~$R$,
$\tr\{T^aT^b\}=T(R)\delta_{ab}$, which is positive-definite [i.e., $T(R)>0$]
for any non-trivial representation~$R$ of the compact group~$G$. In
deriving this expression, we have assumed that the regularization respects
covariance under Lorentz and global $G$ transformations. The scalar function
$L(p^2)$ depends on the regularization applied to the diagram. For example, the
Pauli-Villars regularization respects the covariance and gives $L(p^2)=-1$. As
we see below, however, the existence of a massless pole is independent
of the precise form of~$L(p^2)$, and thus of the regularization. The current
divergence is therefore given by
\begin{eqnarray}
   &&\mathop{\rm FT}\langle0|
   T^*\partial_\mu\{\overline\psi\gamma^\mu\gamma_5T^a\psi(x)\}
   \overline\psi\gamma_\nu T^b\psi(y)|0\rangle
\nonumber\\
   &&\qquad\qquad\qquad\qquad\qquad\qquad
   =-{1\over2\pi}(\dim r)T(R)\delta_{ab}
   \{1-L(p^2)\}\epsilon_{\nu\rho}p^\rho,
\nonumber\\
   &&\mathop{\rm FT}\langle0|
   T^*\overline\psi\gamma_\mu\gamma_5T^a\psi(x)
   \partial_\nu\{\overline\psi\gamma^\nu T^b\psi(y)\}|0\rangle
\nonumber\\
   &&\qquad\qquad\qquad\qquad\qquad\qquad
   ={1\over2\pi}(\dim r)T(R)\delta_{ab}
   \{1+L(p^2)\}\epsilon_{\mu\rho}p^\rho.
\label{twoxnine}
\end{eqnarray}
We thus conclude that Eq.~(\ref{twoxseven}) is, to all orders in the
perturbation theory, given by the expressions in Eq.~(\ref{twoxnine}).
Explicitly, we have $F^{ab}(p^2)=(\dim r)T(R)\delta_{ab}$ and
$G^{ab}(p^2)=(\dim r)T(R)\delta_{ab}L(p^2)$, and hence we have
\begin{eqnarray}
   &&\mathop{\rm FT}\langle0|
   T^*J_{5\mu}^a(x)J_\nu^b(y)|0\rangle
\nonumber\\
   &&=-{i\over2\pi}(\dim r)T(R)\delta_{ab}
   \left\{
   {1\over p^2}(p_\mu\epsilon_{\nu\rho}p^\rho
   +p_\nu\epsilon_{\mu\rho}p^\rho)+L(p^2)\epsilon_{\mu\nu}
   \right\}
\label{twoxten}
\end{eqnarray}
as the expression which holds to all orders in perturbation theory. It is
surprising that the two-point function is completely determined by the one-loop
diagram and receives no higher-order radiative corrections; this is a
consequence of the (anomalous) WT identities. In other words, we showed that
the level of the Kac-Moody algebra does not receive higher-order corrections
in the present super-renormalizable theories. One can also explicitly confirm
that there are no radiative corrections to the two-point function, for example
at the two-loop order.

From the expression~(\ref{twoxten}), we observe that there exists a massless
pole which corresponds to massless bosonic states (since $J_{5\mu}^a$ and
$J_\nu^b$ are bosonic operators).\footnote{Here we assume that the vacuum is
a bosonic state.} This pole cannot be eliminated by the regularization
ambiguity~$L(p^2)$, because the first term in Eq.~(\ref{twoxten}) is symmetric
under the exchange $\mu\leftrightarrow\nu$, while the second is anti-symmetric.
Recall also that $T(R)>0$ for any (non-trivial) representation~$R$. Thus, the
residue is always non-zero for any (non-trivial) fermion content.

We note that the existence of massless bosonic states does not contradict
Coleman's theorem,\cite{Coleman:1973ci} which rules out the spontaneous
breaking of bosonic symmetry in two dimensions. The spontaneous symmetry
breaking implies that a massless bosonic state, the Nambu-Goldstone boson,
appears in the intermediate state in a channel between a conserved current and
a scalar field. It can be shown that,\cite{Coleman:1973ci} in two dimensions,
such a massless intermediate state is inconsistent with the normalizability
and positivity of physical states.\footnote{An example demonstrating this
statement is
provided by the free massless scalar field~$\phi(x)$ in two dimensions. In this
system, the shift symmetry $\phi(x)\to\phi(x)+\epsilon$ of the action
{\it is\/} always spontaneously broken, because the vacuum expectation value of
$i[Q,\phi(x)]=1$, where the charge $Q=\int\rmd x^1\,j^0(x)$ is defined in terms
of the
conserved current $j_\mu(x)=\partial_\mu\phi(x)$, is unity. Correspondingly, in
the intermediate state between $j_\mu$ and $\phi$, there appears the massless
Nambu-Goldstone boson that is simply $\phi$ itself. However, the state
created by $\phi(x)$ from the vacuum is not normalizable, because $\delta(p^2)$
has no well-defined Fourier transform in two dimensions (due to the
infrared divergence). Spontaneous symmetry breaking does occur in this theory,
but this theory itself is ill-defined.} This argument, however, does not place
any restriction on possible massless states appearing in a channel between
{\it two conserved currents}, such as Eq.~(\ref{twoxten}). In fact, the
massless pole in Eq.~(\ref{twoxten}) has nothing to do with the spontaneous
symmetry breaking. Rather, it corresponds to (as we have already noted) the
central extension of the current algebra. These points can be explicitly seen
by applying the
Bjorken-Johnson-Low (BJL) prescription\cite{Bjorken:1966jh,Johnson:1966se} to
Eq.~(\ref{twoxten}). Assuming that the function~$L(p^2)$ is a polynomial in
$p_0$, we have (see Appendix~A)
\begin{equation}
   \langle0|[J_5^{a0}(x),J_\nu^b(y)]|0\rangle\delta(x^0-y^0)
   =-{i\over\pi}(\dim r)T(R)\delta_{ab}\delta_\nu^0
   \delta(x^0-y^0){\rmd\over\rmd x^1}\delta(x^1-y^1).
\label{twoxeleven}
\end{equation}
Integrating both sides over the spatial coordinate $x^1$, we see that the
vacuum expectation value of the commutator between the axial charge
$Q_5^a=\int\rmd x^1\,J_5^{a0}(x)$ and the vector current~$J_\nu^b(y)$ vanishes.
This clearly demonstrates that the existence of the massless pole does not
imply the spontaneous
symmetry breaking, and also that the $G$-algebra, $[Q_5^a,Q^b]=if_{abc}Q_5^c$,
where $Q^b=\int\rmd x^1\,J^{b0}(x)$ is the vector charge, does not suffer from
the ($c$-number) anomaly. It is interesting that in the present two-dimensional
models, the situation is the opposite of that described in the Nambu-Goldstone
theorem: The
$G_L\times G_R$ symmetry remains exact, {\it and\/} there arise massless bosonic
states.

Equation~(\ref{twoxten}) implies that there exists a state $|X\rangle$ such
that\footnote{More specifically, when $Y=0$,
Eq.~(\ref{twoxten}) shows that there exists
a massless asymptotic field~$\tilde J_\mu^a(x)$ that can be expanded in the
creation and annihilation operators as
\begin{equation}
   J_\mu^a(x)\to Z^{1/2}\tilde J_\mu^a(x)=
   Z^{1/2}\int{\rmd k_1\over\sqrt{2\pi\cdot2k_0}}\left\{
     -ik_\mu a^a(k_1)e^{-ikx}+ik_\mu a^{a\dagger}(k_1)e^{ikx}\right\}
   \qquad\hbox{for $x^0\to+\infty$},
\end{equation}
where $k_0=|k_1|$ and $Z=(1/\pi)(\dim r)T(R)$. When acting on the
vacuum~$|0\rangle$, the asymptotic field $\tilde J_\mu^a(x)$ and the
creation operator~$a^{a\dagger}(k_1)$ define {\it normalizable\/} states. Note
that this is in sharp contrast to the case of a massless asymptotic scalar
field $\tilde\phi(x)=\int{\rmd k_1\over\sqrt{2\pi\cdot2k_0}}
\{a(k_1)e^{-ikx}+a^\dagger(k_1)e^{ikx}\}$, for which the
state~$\tilde\phi(x)|0\rangle$ is {\it not\/} normalizable due to the
infrared divergence caused by the factor~$\sqrt{k_0}$ in the denominator. In
the case considered presently, therefore, the LSZ reduction formula provides
well-defined transition amplitudes when applied to multi-point functions of
current operators. For example, from a fermion one-loop diagram which contains
three vector currents, we have
\begin{eqnarray}
   &&\int\rmd^2x\,e^{ipx}\int\rmd^2y\,e^{iqy}\,
   \langle k,c|T^*J_\mu^a(x)J_\nu^b(y)|0\rangle
\nonumber\\
   &&={-2\pi i\over\sqrt{2\pi\cdot2k_0}}\,Z^{-1/2}
   (\dim r)T(R)f^{abc}\delta^2(p+q+k)
\nonumber\\
   &&\quad\times
   {1\over p^2-q^2}\left(\left\{
   (p_\mu-q_\mu)k_\nu+(p_\nu-q_\nu)k_\mu+g_{\mu\nu}(p^2-q^2)
   +2k_\mu k_\nu\,{p^2+q^2\over p^2-q^2}\right\}
   \ln\left\{{p^2\over q^2}\right\}
   -4k_\mu k_\nu\right),
\end{eqnarray}
where $\langle k,c|\equiv\langle0|a^c(k_1)$. This expression of a well-defined
transition amplitude, which holds at zeroth order in gauge and Yukawa
couplings, illustrates that these massless asymptotic states pose no problem
in defining the S matrix through the LSZ formula.}
\begin{equation}
   \langle X|J_\mu^a(0)|0\rangle\neq0.
\end{equation}
Assuming covariance under global $G$ transformations, we see that the massless
state belongs to a non-trivial multiplet of $G$ [unless $G=\U(1)$] and that the
right-hand side of the above equation can be written in terms of invariant
tensors of $G$. The simplest possibility would be that $|X^b\rangle$ transforms
as a conjugate of the current $J_\mu^b$ under $G$ transformations, and the
right-hand side is the invariant tensor~$\delta_{ab}$. Thus the minimum number
of massless bosonic states {\it per fixed spatial momentum~$p_1$\/} is given by
the index~$a$, the dimension of the group~$G$.

Moreover, the bosonic state $|X\rangle$ is physical in the sense that it can be
chosen to contain no unphysical modes, such as FP ghosts and longitudinal
modes of gauge fields. This can be seen most clearly by using the form of the
completeness relation in the present gauge system, $1=P^{(0)}+\{Q_B,R\}$. Here
$P^{(0)}$ is the projection operator to the Hilbert
space~${\mathcal H}_{\rm phys}$ that does not contain any unphysical modes and
$Q_B$ is the BRST
charge.\cite{Kugo:1979gm} \ The second term in the completeness relation, where
$R$ is a certain operator, is the projection operator to states which contain
at least one unphysical mode. Since the global currents $J_{5\mu}^a$ and
$J_\nu^b$ are gauge invariant and commute with the BRST charge, the second term
in the completeness relation does not contribute when inserted into the
two-point
function~(\ref{twoxten}). This shows that the state $|X\rangle$ is an element
of ${\mathcal H}_{\rm phys}$.\footnote{Since there is ambiguity in adding BRST
exact states to $|X\rangle$, it is more precise to say that $|X\rangle$ can be
chosen as an element of ${\mathcal H}_{\rm phys}$.}

We must be careful, however, in the interpretation of the massless bosonic
states. We have intentionally used the term ``state'' instead of ``particle'',
because in two dimensions, even multi-particle intermediate states can produce
a massless pole. This point becomes clear if we consider a system of free
massless fermions (see the next section). It is thus not obvious whether the
massless pole we observed corresponds to a single massless boson or is due to
(say) two massless fermions. In two dimensions, it is not clear which
interpretation is appropriate, as the bosonization of fermions is possible.
Nevertheless,
Eq.~(\ref{twoxten}) provides non-trivial information regarding the low-energy
spectrum
of the model. In particular, if we assume the validity of Eq.~(\ref{twoxten})
in a non-perturbative level (like the anomaly matching condition), it
constrains the
possible patterns of low-energy spectra with possible assignment of the
$G$-representation.

As a final remark, we explain why $G_L\times G_R$ symmetry should not be gauged
for our argument to be applied. In other words, our argument is not applied
to the two-point function of {\it gauge currents}. The reason is that we use the
fact that the $G_L\times G_R$ symmetry is chiral (i.e., it involves
$\gamma_5$). Gauging $G_L\times G_R$ means that we introduce two distinct gauge
fields, one coupled to $J_{L,\mu}^a$ and the other coupled to
$J_{R,\mu}^a$; the resulting system is a chiral gauge theory. In two
dimensions, however, such a chiral gauge theory is always anomalous, and it
would be meaningless to consider such a theory from the outset. The best thing
we can do is to gauge the vector sub-group $G_V$ of $G_L\times G_R$. In such
a system, however, there are an infinite number of potentially UV diverging
diagrams (chains of fermion one-loop diagrams connected by gauge propagators)
which contribute to the two-point function of currents. This infinite set of
diagrams can produce a massive pole, as is well known for the two-point
function of currents in the single-flavor Schwinger model.
It is thus seen that the
assumption that $G_L\times G_R$ is not gauged is important for our argument.

In the next section, we first illustrate the power of our argument by
considering exactly solvable models. Then we consider its application to
non-solvable models, two-dimensional supersymmetric gauge theories.

\section{Applications}
\subsection{Solvable models}
The simplest example to which the above argument can be applied is a system
of $n$~free massless Dirac fermions. The global symmetry in this system is
$\U(n)_L\times\U(n)_R$, and hence we expect the existence of at least $n^2$
massless bosonic states (per fixed spatial momentum). As is well known, this
system can equivalently be expressed by $n$~free massless real scalar
fields---the so-called abelian
bosonization.\cite{Coleman:1974bu,Mandelstam:1975hb,Halpern:1975nm} \ The
existence of massless bosonic states itself is thus manifest in this picture of
bosonized theory.

In the abelian bosonization, among the global currents $J_{L,\mu}^a$
and~$J_{R,\mu}^a$, those associated with the Cartan sub-algebra of $\U(n)$ are
simply given by derivatives of the scalar fields, and the appearance of a
massless pole in the corresponding two-point functions is obvious. Other
currents, associated with off-diagonal generators of $\SU(n)$, are non-local
and non-polynomial functions of the scalar fields, and therefore the appearance
of a massless pole is less obvious. This asymmetric treatment of currents is
necessary for the abelian bosonization, and a symmetric treatment is possible
through non-abelian bosonization,\cite{Witten:1983ar} which uses $n^2$~real
scalar fields. In any case, the two-point functions are given by
Eq.~(\ref{twoxten}), which indicates (at least) $n^2$ massless states.

This example of a free fermion illustrates subtlety in the interpretation of
the massless pole in~Eq.~(\ref{twoxten}) in terms of particles. In the original
theory of fermions, which corresponds precisely to the diagram appearing in
Fig.~\ref{fig4}, a pair of massless fermions produces the massless pole. In the
bosonized theory, the pole is produced by the exchange of massless bosons.
These pictures are equivalent.

The first non-trivial example is two-dimensional QED with $n$~massless
electrons (where $n>1$), i.e., the massless Schwinger
model\cite{Schwinger:1962tp,Lowenstein:1971fc} with many
flavors.\cite{Belvedere:1978fj} \ In this case, the global symmetry that is not
gauged and
does not suffer from the anomaly is $\SU(n)_L\times\SU(n)_R$, and thus we
expect at least $n^2-1$ massless bosonic states. As emphasized in
Ref.~\citen{Halpern:1975jc}, the global currents $J_{L,\mu}^a$
and~$J_{R,\mu}^a$
associated with the $\SU(n)_L\times\SU(n)_R$ symmetry can be expressed in terms
of $n-1$~free massless real scalar fields. The existence of massless bosonic
states itself is manifest in this bosonized theory, and the
expression~(\ref{twoxten}), which indicates (at least) $n^2-1$ massless states,
is reproduced. It is interesting that the single-flavor Schwinger
model\cite{Schwinger:1962tp,Lowenstein:1971fc} has no (physical)
massless bosonic states. The would-be global symmetry $\U(1)_V\times\U(1)_A$ in
this model either is gauged or suffers from the anomaly; hence our argument
does not apply.

A somewhat different type of solvable non-gauge model is the Kogut-Sinclair
model,\cite{Kogut:1975iz,Witten:1978qu} in which the Yukawa interaction takes
the form
\begin{equation}
   Y=\overline\psi e^{i\gamma_5\pi}\psi,
\end{equation}
and the global $\U(1)_L\times\U(1)_R$ symmetry is realized by
Eq.~(\ref{twoxtwo}) with $\pi\to\pi+\theta_L$ and Eq.~(\ref{twoxthree})
with $\pi\to\pi-\theta_R$. (There is no scalar potential~$V$ in this model.)
The abbreviated term of $J_{5\mu}^a$ in Eq.~(\ref{twoxfive}) is linear in the
pseudo-scalar~$\pi$. Again, according to the previous argument, we conjecture
the existence of one
massless bosonic state. Indeed, here, this is the case, as seen from the exact
solution.\cite{Kogut:1975iz,Witten:1978qu} What is more interesting about this
model is that one can observe saturation of the residue of the massless pole by
a massless boson and the absence of fermions with a non-zero $\U(1)_A$
charge.\cite{Kogut:1975iz,Witten:1978qu} This picture provides a very
non-trivial realization of Eq.~(\ref{twoxten}), because the elementary fermion
in the original lagrangian has a non-zero $\U(1)_A$ charge.

The two-dimensional $\SU(N)$ gauge theory with $n$~massless fermions (i.e.,
two-di\-men\-sion\-al massless QCD) is not exactly solvable. Nevertheless, it
can be shown that this model contains $n^2$ massless bosonic
states\cite{Elitzur:1981gh,Buchmuller:1981ga} for finite~$N$, as well as for
$N\to\infty$.\cite{'tHooft:1974hx} \ (See also
refs.~\citen{Polyakov:1988qz} and \citen{Kutasov:1994xq}.)
This is in accord with the above argument,
because the global symmetry of this model is $\U(n)_L\times\U(n)_R$. We note
that our counting of the minimal number of massless bosonic states is
independent of the gauge-group representation of fermions.

\subsection{Two-dimensional supersymmetric gauge theories}
A particularly interesting application is provided by supersymmetric gauge
theories in two dimensions, which are not exactly solvable. We are thus
interested in their spectra, especially at low energies. As the parity
invariant two-dimensional supersymmetric gauge theories, we have
$\mathcal{N}=(1,1)$, $(2,2)$, $(4,4)$ and $(8,8)$ models. In the
$\mathcal{N}=(1,1)$ supersymmetric gauge theory, the gaugino is a single
Majorana fermion, and consequently the action generically does not possess any
continuous symmetry of the $G_L\times G_R$ type. Here we do not attempt
to carry out a general
classification of the $\mathcal{N}=(1,1)$ models. Rather, as a special case,
we first consider the $\mathcal{N}=(2,2)$ supersymmetric $\SU(N)$ pure
Yang-Mills (YM) theory.

In the $\mathcal{N}=(2,2)$ $\SU(N)$ pure YM theory, the Yukawa interaction and
the scalar potential take the forms
\begin{equation}
   Y=g\tr\left\{\overline\psi\left[\sigma+i\gamma_5\pi,\psi\right]\right\},
   \qquad
   V=-g^2\tr\left\{\left[\sigma,\pi\right]^2\right\},
\end{equation}
where all fields belong to the adjoint representation of the gauge
group~$\SU(N)$, and the commutator and the trace are taken with respect to this
gauge-group representation. From these, we see that the model possesses
$\U(1)_L\times\U(1)_R$ symmetry and that, applying the above argument, there
must exist at least one massless bosonic state (per fixed spatial momentum).
These conclusions are valid even in the case that there exist matter
multiplets, as long as there is no superpotential. The two-dimensional
$\mathcal{N}=(2,2)$
supersymmetric gauge theory can be obtained from the four-dimensional
$\mathcal{N}=1$ supersymmetric gauge theory by dimensional reduction. The
rotational $\SO(2)$ invariance in the reduced dimensions becomes $\U(1)_A$ in
two dimensions, and thus the two-dimensional $\mathcal{N}=(2,2)$ models always
possess the $\U(1)_A$ symmetry with $\U(1)_A$ charges (i.e., the
representation) fixed by the underlying rotational invariance. The $\U(1)_A$
charges come to be unity for all fermions. Then, the $\U(1)_V$ charges for each
fermion are fixed (to unity), because they must be identical to the $\U(1)_A$
charges for our argument to apply. It turns out that any superpotential is
inconsistent with this assignment of the $\U(1)_V$ charges.

If the gauge group contains the $\U(1)$ factor, like the $\mathcal{N}=(2,2)$
supersymmetric massless QED, we need  ``anomaly cancellation'' for the current
conservation of global currents $J_{L,\mu}$ and~$J_{R,\mu}$, as explained in
Fig.~1. This requires that, for each $\U(1)$ factor, the sum of the $\U(1)$
charges~$Q_i$ of the fermion over all matter multiplets vanishes, i.e.,
$\sum_iQ_i=0$. If this condition is satisfied (and there is no superpotential),
there must exist at least one massless bosonic state in this
system.\footnote{The possible presence of the Fayet-Iliopoulos $D$-term does
not change the conclusion, because after the integration over auxiliary fields,
the lagrangian density takes the form of Eq.~(\ref{twoxone}). However,
we cannot include the $\theta$-term in the present argument, because it
breaks the parity invariance.}

If we consider the $\mathcal{N}=(4,4)$ supersymmetric pure YM theory, which is
a special case of the $\mathcal{N}=(2,2)$ supersymmetric gauge theory without
a superpotential, the symmetry is enhanced to $\SU(2)_L\times\SU(2)_R$. In this
case, the
Yukawa interaction and the scalar potential of the $\mathcal{N}=(4,4)$ pure YM
theory are given by
\begin{eqnarray}
   &&Y=g\tr\left\{\overline\psi
   \left[\sigma+i\gamma_5\pi_i\tau^i,\psi\right]
   \right\},
\nonumber\\
   &&V=-g^2\tr\left\{\left[\sigma,\pi_i\right]^2
   +\left({1\over2}\epsilon_{ijk}\left[\pi_j,\pi_k\right]\right)^2\right\},
\end{eqnarray}
where the pseudo-scalars $\pi_i$ ($i=1$, 2, 3) and fermions~$\psi$ and
$\overline\psi$ are the triplet and doublets of the flavor
$\SU(2)$, respectively, and $\tau^i$ denotes the Pauli matrices. The theory thus
possesses $\SU(2)_L\times\SU(2)_R$ global symmetry, which is a realization of
the $\SO(4)$ part of the $R$~symmetry of this model. According to the above
argument, therefore, there must be at least three massless bosonic states (per
fixed spatial momentum).

As soon as we couple matter multiplets to the $\mathcal{N}=(4,4)$ pure YM
theory, however, all chiral symmetries are broken, because the
$\mathcal{N}=(4,4)$ supersymmetry requires that we introduce a particular form
of the superpotential [in terms of the $\mathcal{N}=(2,2)$ theory]. As already
noted, then, the lagrangian has no $G_L\times G_R$ symmetry. Since
the $\mathcal{N}=(8,8)$ gauge theory is a particular case of the
$\mathcal{N}=(4,4)$ gauge theory with matter multiplets, the
$\mathcal{N}=(8,8)$ theory also possess no $G_L\times G_R$ symmetry,
and again our argument does not apply.

Thus we have observed that the $\mathcal{N}=(2,2)$ models without a
superpotential and the $\mathcal{N}=(4,4)$ pure YM theory contain massless
bosonic states. The fact that the $\mathcal{N}=(2,2)$ and
$\mathcal{N}=(4,4)$ pure YM theories have no mass gap was noted in
Ref.~\citen{Witten:1995im} on the basis of the 't~Hooft anomaly matching
condition.

In supersymmetric theories, if supersymmetry is not spontaneously
broken, a massless bosonic state implies a massless fermionic state. For
example, in the $\mathcal{N}=(2,2)$ pure YM theory, Eq.~(\ref{twoxten})
and a supersymmetric WT identity show that there exists a massless fermionic
state which is created by the action of the supercurrent on the
vacuum.\footnote{This occurs, however, in a way that does not directly imply
spontaneous breaking of supersymmetry.} This fact provides a further constraint
on possible low-energy spectra and should be very useful in the examination of
recently developed lattice formulations of two-dimensional supersymmetric gauge
theories.\cite{Kaplan:2002wv}\tocite{Endres:2006ic} \ The details of such a
study will be reported elsewhere.\cite{We}

\section{Conclusion}
In summary, using an elementary argument, we showed that in a wide class of
$G_L\times G_R$ invariant two-dimensional super-renormalizable field theories,
the parity-odd part of the two-point function of global currents can be
determined to all orders in perturbation theory. For any non-trivial fermion
content, the two-point function possesses a massless pole, and this fact
provides a simple criterion for the existence of massless bosonic states in the
theory.
As a particular application, we considered two-dimensional supersymmetric gauge
theories.

\section*{Acknowledgements}
The authors thank the member of the Yukawa Institute for Theoretical Physics at
Kyoto
University, where this work was initiated during the YITP workshop on ``Actions
and symmetries in lattice gauge theory'' (YITP-W-05-25). We would like to thank
the participants in the workshop, especially Poul Damgaard, Peter Hasenfratz,
David B. Kaplan and Yoshio Kikukawa for suggestions. We also thank So Matsuura
and Kazutoshi Ohta for discussions. This work is supported in part by MEXT
Grants-in-Aid for Scientific Research, Nos.~13135203, 18540305 (H.~S.)
13135223 and~18034011 (M.~H.), and by JSPS and the French Ministry of Foreign
Affairs under the Japan-France Integrated Action Program (SAKURA). I.~K. is
supported by the Special Postdoctoral Researchers Program at RIKEN.

\appendix
\section{BJL prescription and the derivation of Eq.~(\protect\ref{twoxeleven})}
The BJL prescription\cite{Bjorken:1966jh,Johnson:1966se} enables one to
extract the $T$~product from the corresponding $T^*$~product. (See also
Ref.~\citen{Fujikawa:1986bs} for an analysis of a related problem.) The
prescription (for two-point functions) is based on the
following two requirements:
(1)~The possible difference between the $T$~product and the $T^*$~product
arises only at equal times. (2)~The $p_0\to\infty$ limit of the Fourier
transform of the $T$~product vanishes.\cite{Bjorken:1966jh} From these, the
Fourier transform of the $T$~product is obtained from that of the
corresponding $T^*$~product by subtracting a {\it polynomial\/} of $p_0$, so
that the $p_0\to\infty$ limit vanishes.

For example, from Eq.~(\ref{twoxten}), we have
\begin{eqnarray}
   &&\mathop{\rm FT}\langle0|TJ_{5\mu}^a(x)J_\nu^b(y)|0\rangle
\nonumber\\
   &&=-{i\over2\pi}(\dim r)T(R)\delta_{ab}\left\{
   {1\over p^2}(p_\mu\epsilon_{\nu\rho}p^\rho
   +p_\nu\epsilon_{\mu\rho}p^\rho)
   -\delta_\mu^0\epsilon_{\nu0}-\delta_\nu^0\epsilon_{\mu0}\right\}.
\label{axone}
\end{eqnarray}
Similarly, from the divergence of Eq.~(\ref{twoxten}), we have
\begin{equation}
   \mathop{\rm FT}
   \langle0|T\partial_\mu J_5^{a\mu}(x)J_\nu^b(y)|0\rangle=0.
\label{axtwo}
\end{equation}
Now, the divergence of Eq.~(\ref{axone}) yields
\begin{equation}
   \mathop{\rm FT}\partial_\mu
   \langle0|TJ_5^{a\mu}(x)J_\nu^b(y)|0\rangle
   =-{1\over\pi}(\dim r)T(R)\delta_{ab}\delta_\nu^0p_1,
\label{axthree}
\end{equation}
which is, using Eq.~(\ref{axtwo}), precisely the commutator
$\mathop{\rm FT}\langle0|[J_5^{a0}(x),J_\nu^b(y)]|0\rangle\*\delta(x^0-y^0)$,
and then we have Eq.~(\ref{twoxeleven}).

\section{Derivation of Eqs.~(2.12) and~(2.13)}
When $Y=0$, we can set $J_{5\mu}^a(x)=-\epsilon_{\mu\nu}J^{a\nu}(x)$ and
Eq.~(\ref{twoxten}) implies
\begin{equation}
   \mathop{\rm FT}\langle0|
   T^*J_\mu^a(x)J_\nu^b(y)|0\rangle
   ={i\over2\pi}(\dim r)T(R)\delta_{ab}
   \left\{
   {2p_\mu p_\nu\over p^2}+(L(p^2)-1)g_{\mu\nu}
   \right\}.
\label{bxone}
\end{equation}
Then the asymptotic field for the vector current $Z^{1/2}\tilde J_\mu^a(x)$ can
be obtained as a free field whose two-point function reproduces the pole part
of the above two-point function
\begin{equation}
   \mathop{\rm FT}\langle0|
   T^*J_\mu^a(x)J_\nu^b(y)|0\rangle\Bigr|_{\rm pole\ part}
   ={i\over\pi}(\dim r)T(R)\delta_{ab}
   {p_\mu p_\nu\over p^2}.
\label{bxtwo}
\end{equation}
It is thus given by Eq.~(2.12) in which the creation and annihilation operators
satisfy the commutation relation
\begin{equation}
   [a^a(k_1),a^{b\dagger}(q_1)]=\delta_{ab}\delta(k_1-q_1).
\label{bxthree}
\end{equation}
If we define the polarization vector by
\begin{equation}
   \varepsilon_\mu^{(s)}(k_1)={i(|k_1|,-k_1)\over2|k_1|^2},
\label{bxfour}
\end{equation}
the polarization vector possesses a non-zero inner product with the on
mass-shell momentum $k_\mu=(|k_1|,k_1)$,
\begin{equation}
   \varepsilon^{(s)*\mu}(k_1)(-ik_\mu)=-1,
\label{bxfive}
\end{equation}
and the annihilation operator can be extracted from the asymptotic field by
\begin{equation}
   a^a(k_1)=-i\int\rmd x_1\,\varepsilon^{(s)*\mu}(k_1)
   (e^{ikx}\overleftrightarrow{\partial_0}\tilde J_\mu^a(x))
   /\sqrt{2\pi\cdot2k_0}.
\label{bxsix}
\end{equation}
We can then repeat the standard argument for the LSZ formula and we have,
for example,
\begin{eqnarray}
   &&\langle0|a^c(k_1)T^*J_\mu^a(x)J_\nu^b(y)|0\rangle
\nonumber\\
   &&=Z^{-1/2}
   \int\rmd^2z\,{\varepsilon^{(s)*\rho}(k_1)\,e^{ikz}\over\sqrt{2\pi\cdot2k_0}}
   (-i\overrightarrow\square_z)
   \langle0|T^*J_\mu^a(x)J_\nu^b(y)J_\rho^c(z)|0\rangle.
\label{bxseven}
\end{eqnarray}
On the other hand, a direct calculation of fermion one-loop diagrams yields
\begin{eqnarray}
   &&\langle0|T^*J_\mu^a(x)J_\nu^b(y)J_\rho^c(z)|0\rangle
\nonumber\\
   &&\sim-{i\over2\pi}(\dim r)T(R)f^{abc}
   \int{\rmd^2p\over(2\pi)^2}\int{\rmd^2q\over(2\pi)^2}\,
   e^{-ip(x-y)-iq(x-z)}{1\over q^2}
\nonumber\\
   &&\qquad
   \times{q_\rho\over p\cdot q}\biggl\{
   (p_\mu q_\nu+q_\mu p_\nu-g_{\mu\nu}p\cdot q)
   \ln\left\{{p^2+2p\cdot q\over p^2}\right\}
\nonumber\\
   &&\qquad\qquad\qquad\qquad\qquad
   +2q_\mu q_\nu\left(1-{p^2\over2p\cdot q}
   \ln\left\{{p^2+2p\cdot q\over p^2}\right\}
   \right)
   \biggr\},
\label{bxeight}
\end{eqnarray}
where only the terms which possess a massless pole with respect to the
momentum~$q$ have been retained (other terms do not contribute to the
transition amplitude~(2.13) because of the operator $\overrightarrow\square_z$
in Eq.~(\ref{bxseven})). Plugging Eq.~(\ref{bxeight}) into Eq.~(\ref{bxseven}),
we have the transition amplitude~(2.13), where the expression has been somewhat
simplified by using the momentum conservation and the mass-shell condition
$k^2=0$.

\end{document}